\documentclass[12pt]{article}

\usepackage{graphicx}
\usepackage{url}
\usepackage{cite}
\usepackage{amssymb}
\usepackage{amsmath}
\usepackage{placeins}
\usepackage{longtable}
\usepackage[margin=1in]{geometry}
\usepackage{setspace}
\usepackage{array}
\newcolumntype{P}[1]{>{\raggedright\arraybackslash}p{#1}}
\usepackage{ragged2e}
\usepackage{longtable, array} 

\usepackage[margin=1in]{geometry}
\usepackage{setspace}
\usepackage{authblk} 
\usepackage{hyperref}

\title{An Overview of the Prospects and Challenges of Using Artificial Intelligence for Energy Management Systems in Microgrids}

\author[1]{Noor ul Misbah Khanum}
\author[1]{Hayssam Dahrouj}
\author[1,2]{Ramesh C. Bansal}
\author[1]{Hissam Mouayad Tawfik}

\affil[1]{Department of Electrical Engineering, University of Sharjah, Sharjah, United Arab Emirates}
\affil[2]{Department of Electric, Electronic and Computer Engineering, University of Pretoria, Pretoria, South Africa}

\date{12 May 2025}
\begin{document}

\maketitle






%
\begin{abstract}
Microgrids have emerged as a pivotal solution in the quest for a sustainable and energy-efficient future. While microgrids offer numerous advantages, they are also prone to issues related to reliably forecasting renewable energy demand and production, protecting against cyberattacks, controlling operational costs, optimizing power flow, and regulating the performance of energy management systems (EMS). Tackling these energy management challenges is essential to facilitate microgrid applications and seamlessly incorporate renewable energy resources. Artificial intelligence (AI) has recently demonstrated immense potential for optimizing energy management in microgrids, providing efficient and reliable solutions. This paper highlights the combined benefits of enabling AI-based methodologies in the energy management systems of microgrids by examining the applicability and efficiency of AI-based EMS in achieving specific technical and economic objectives. The prospects of such objectives, as illustrated in the paper, include enhancing energy efficiency, demand management, reducing operational costs, improving forecasting and predictive maintenance, and enhancing microgrid resilience and cybersecurity. In this regard, the paper provides promising insights into various prospects that showcase the cost and operational resilience advantages of AI-based EMS. Additionally, the paper depicts the challenges that must be addressed to meet such prospects, namely, data quality and availability, interoperability and integration, scalability and operational challenges, and standardization and regulatory issues. The paper lastly points out several future research directions that promise to spearhead AI-driven EMS, namely the development of self-healing microgrids, integration with blockchain technology, use of Internet of things (IoT), and addressing interpretability, data privacy, scalability, and the prospects to generative AI in the context of future AI-based EMS.
\end{abstract}

\vspace{1em}



\section{Introduction}
\label{sec:introduction}

\subsection{Overview}
\label{subsec1}
The effect of climate change is extensive, and carbon emissions produced by burning fuel to generate electricity have an exacerbating impact on the environment~\cite{b1}. Microgrids incorporate renewable energy resources, energy storage systems, and combined heat power units (CHPs) along with the main grid network, where renewable energy sources play a key role in managing the impacts of climate change, as they utilize clean energy to generate power. With increasing global efforts to reduce carbon footprints, renewable energy sources are being widely adopted, necessitating efficient energy management systems (EMS) to handle the variability and uncertainty of these sources. Given the active global initiatives to achieve net zero carbon emissions by 2050 as envisioned by the United Nations through its Division of Sustainable Development Goals (DSDG), it becomes crucial to highlight the prospects, challenges, and techniques to develop the next generations of EMS~\cite{b2}. Artificial intelligence is nowadays transforming energy management systems by changing the way energy is generated, distributed, and consumed. Therefore, integrating AI-based solutions is bound to achieve more resilient, reliable, and responsive EMS. The motivation behind this review paper is to explore how AI can potentially overcome the limitations of conventional EMS and explore new capabilities for future microgrids. This paper focuses on the potential advantages and technical challenges offered by the integration of Artificial Intelligence (AI) tools in designing the next generations of EMS in future microgrids. The paper first starts by presenting the conventional control system of microgrids and their energy management, along with the basics of AI tools and techniques. Then, the features and potential advantages of AI-based EMS against conventional energy management systems in microgrids are highlighted.

\subsection{Microgrids and their energy management system}
Recently, there has been an increasing growth in the incorporation of renewable energy resources (RERs) into the power distribution grid~\cite{b3}. Hybrid renewable energy systems, which combine multiple clean energy sources, are becoming increasingly popular to streamline the advancement of the next generation of power systems. Microgrids have emerged as a possible solution to reduce the growing energy demands~\cite{b4}. Microgrids also contribute to a sustainable and resilient energy infrastructure. Renewable energy resources, energy storage systems like batteries, and CHPs together make up the distributed energy resources (DERs)~\cite{b5}. A local power grid that is self-reliant is called a microgrid, and consists of several DERs.

The literature on microgrids distinguishes between two types of microgrids, namely, grid-connected and off-grid~\cite{b6}. In the grid-connected microgrid, the utility grid is connected to the microgrid and helps provide extra energy to the main grid. In the off-grid type, however,  the microgrid runs independently without any interference from the main utility grid. The off-grid type distributes electricity to local and small communities and also helps to overcome the challenges resulting from the power outage of the main grid~\cite{b7}. The two types of microgrids are not mutually exclusive and can operate within the same system.

Microgrids deploy a three-level control system as outlined in~\cite{b8}. The primary layer of control in the first level is responsible for maintaining overall stability by regulating voltage and frequency levels. The second layer of control performs the function of energy management to ensure optimized performance by maintaining effective coordination among microgrid components. The tertiary control layer of a microgrid is responsible for coordinating and communicating with other microgrids. Given the growing interest in automating the energy management modules of microgrids, this paper focuses on the second control layer, the EMS, which manages the energy flows within the microgrid. 

The functionality of microgrids is extended to include more advanced applications that allow for enhanced performance, efficiency, and reliability~\cite{b9}. These enhanced microgrids are known as smart microgrids. Smart microgrids make use of advanced monitoring and control technologies such as sensors, smart meters, and automated control systems to play a crucial role in optimizing and streamlining the sharing and consumption of energy amongst multiple sectors. Traditional smart microgrids are characterized by standard levels of rules and automation, lacking in their predictive and adaptive abilities, which restricts their capacity to manage energy assets and respond to uncertainties~\cite{b10}. Conversely, AI-based microgrids differ from traditional smart microgrids in terms of higher degree of intelligence, dynamic thinking, and optimization capabilities. AI-based microgrids are capable of continuously learning from data, making better energy demand predictions, and dynamically optimizing energy distribution in real-time, thus providing improved efficiency and reliability~\cite{b11}. They have predictive analytics to make proactive corrections and more precise control over the allocation of resources, helping to improve the integration of renewable energy and the functioning of the entire system.  Furthermore, AI-enhanced microgrids offer users detailed insights and recommendations rather than basic monitoring and control as in the case of traditional smart microgrids.

EMS, as a fundamental part of microgrids, addresses challenges related to real-time control, stability, and optimal use of DERs and fulfills certain performance standards. Energy management (EM) involves monitoring, coordinating, and controlling the different stages of the energy dispatch process, from generation to distribution. The EMS enable a microgrid to operate reliably, safely, flexibly, and economically whether it is grid-connected or off-grid~\cite{b12}. The main aim of the EMS is to optimize the number of generator units to maintain an uninterrupted power supply, along with the other key objectives. The EMS can also be viewed from a multi-objective optimization perspective with objectives such as carbon emission, operational cost, power losses, fuel cost, and generation cost~\cite{b13}. Along with the objectives, there are multiple constraints that should be taken care of, such as maximum and minimum generator power, energy storage, and constraints related to economic dispatch. The functions of EMS  are often divided into two types: short-term scheduling, which focuses on real-time economic dispatch depending on the forecast of energy, and long-term scheduling, which utilizes predictions to determine day-ahead requirements for DERs.

\begin{figure}
  \centering
 \includegraphics[width=1\linewidth]{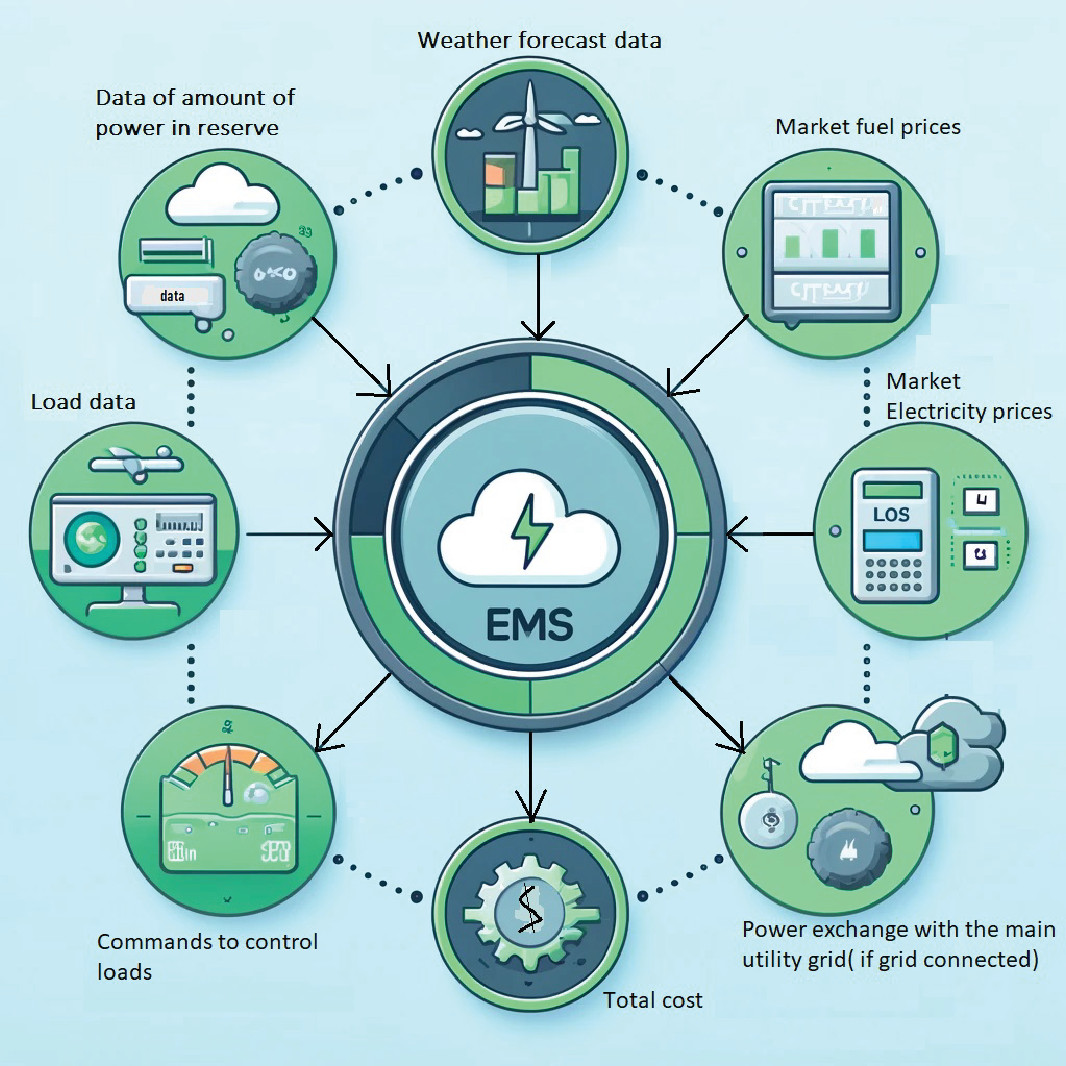}
  \caption{General overview of tasks related to EMS}
   \label{fig: Figure 1}
\end{figure}

Fig. \ref{fig: Figure 1} provides an overview of the EMS, where the input to the EMS is the data related to the amount of power available, forecast of weather and load demand, and market prices for fuel and electricity. To ensure maximum efficiency and avoid delays during peak hours, it is important to track the status of the storage elements present in the microgrid when making decisions related to EM. Data related to the load and weather forecast should be typically obtained 24 hours prior to the scheduling in order to schedule DERs a day ahead and to optimize the power flow in case of any unexpected weather conditions. It is also important to keep track of fuel and electricity prices to reduce the cost. The outputs of EMS are the set points for DERs, generators, and storage elements, which guarantees the optimal functioning of microgrids. Other outputs are the total cost of operation, the exchange of power with the main utility grid if grid-connected, and commands to loads that can be controlled as depicted in Fig.\ref{fig: Figure 1}.  

EMS is designed either in a centralized or decentralized architecture, similar to other control systems. A centralized EMS uses a single microgrid controller that gathers all relevant input data required by the EMS as shown in Fig. \ref{fig: Figure 1}. This controller analyzes the data, finds the best type of action to be taken, and then directs the DER units to make the necessary adjustments, like changing the set points. 

The EMS of microgrids face a number of challenges that affect its efficiency and reliability. For instance, some important challenges are those related to the integration of variable renewable energy sources, the management of systems' complexity, the constraints that arise from limited storage capacities, and effective data management and communication~\cite{b14}. The high costs of implementation and maintenance and the need for scalability and flexibility are also key issues. The cyber risks to system security and the need for user engagement and awareness make the challenges even more alarming. 

Ensuring the reliability and resilience of the EMS, especially during periods of disruption, and the ability to effectively integrate new technologies with the existing infrastructure are other challenges that make efficient energy management a complex task. Effective and efficient energy management is an important challenge that must be addressed, which can eventually help in utilizing microgrids and the incorporation of renewable energy resources. Harnessing the power of AI, energy management in microgrids can be significantly evolved to have a more advanced, efficient, and resilient system. In fact, as power systems become more complex due to advancements in technological and infrastructural developments and the addition of more distributed generators, there is a massive amount of data generated. Regular computer methods struggle to handle and process such data well, and so using AI methods becomes necessary in automating the complex decision-making of future power systems~\cite{b15}.

\subsection{Market drivers and industry needs for AI-based EMS in microgrid}

The global energy landscape is changing rapidly and there is a growing need for sustainable and reliable power solutions. Several key market considerations and industry needs are driving the use of AI in EMS of microgrids. The rapid growth of data centers, especially those supporting AI applications, has significantly increased electricity consumption~\cite{b16,b17} and conventional energy management methods often struggle to manage the complexity and scale of these modern energy demands. AI-based EMS have the potential to optimize energy distribution and consumption in real-time, thereby improving efficiency and reliability. Futhermore, the global shift towards RERs like solar and wind leads to volatility and unpredictability in the grid. There is also an increasing number of power outages and cyber threats, which makes it critical to have resilient and secure energy systems~\cite{b18}. AI enhances microgrid resilience by enabling predictive maintenance and rapid response to attacks. Moreover, industries are under pressure to cut operational costs and meet strict environmental regulations. AI-based EMS are, in fact, poised to help optimizing energy use, leading to decreased costs and ensuring compliance with standards that support integration of RERs. Incorporating AI into microgrid EMS can further address the aforementioned market drivers and industry needs, leading to the development of more efficient, resilient, and sustainable energy systems.

\subsection{The potential of AI-based EMS}

Integration of AI in EMS of microgrids helps in the efficient optimization of energy consumption, enables the integration of renewable sources, and ensures an increase in grid stability. The application of AI algorithms such as machine learning and deep learning has been applied to predict demand patterns in~\cite{b19} and to manage energy storage, and dynamically optimize supply in~\cite{b20} to realize efficient use, minimizing wastage and guarantee availability. AI also integrates intermittent renewable energy sources through a prediction of generation based on weather data and historical trends as analyzed in~\cite{b21}  that accounts for a reduction in the share of fossil fuels and more sustainability.

Moreover, EMS with AI enhances the resilience and economic efficiency of the grid. These systems can undertake predictive maintenance by identifying possible equipment failures before they occur, as analyzed in~\cite{b22}, which helps in minimizing downtimes and associated maintenance costs. The advanced demand response strategies that AI systems carry out continuously adjust loads to the real-time status of the grid to minimize the energy cost and reduce the stress on the grid during peak hours. AI-based systems for market analysis help to make cost-effective decisions by forecasting prices of energy procurement for the electricity markets and energy use optimization is investigated in~\cite{b18}. In general, AI-based EMS for microgrids draw increasing interest due to the increased operational reliability and efficiency, leading to environmental sustainability in the global transition to smart energy systems.

Along these directions, this overview paper provides an overview of the prospects and challenges resulting from
integrating AI tools in microgrid EMS. The main contributions of this paper can be summarized as follows:
\begin{itemize}
    \item The prospects of AI-enabled microgrids are presented in light of energy management by advocating how this integration can help achieving the objectives of enhancing energy efficiency, demand management, and
reducing operational costs, improving forecasting and predictive maintenance, and enhancing microgrid resilience and cybersecurity. 

 \item  The promising insights of AI-enabled microgrid prospects in enhancing energy efficiency, reducing operational costs, and improving microgrid resilience and cybersecurity are illustrated through numerical evaluations.
\item The main challenges that come with the integration of AI in the EMS of microgrids are identified, such as data quality and availability, interoperability and integration, scalability and operational issues, and standardization and regulatory issues.
\item Open research directions for AI-driven EMS in microgrids are discussed which include development of self-healing microgrids, integration with blockchain technology, use of Internet of things, addressing interpretability, data privacy, and scalability and the use of generative AI to achieve more efficiency. 
\end{itemize}

This paper is structured as follows- Section II delves into energy management in microgrids, while Section III provides an overview of AI techniques. Section IV discusses AI-enabled microgrid prospects, followed by the proof of concept using numerical evaluations, which is discussed in Section V. Sections VI and VII present the challenges and open issues faced by integrating AI into the EMS of microgrids, respectively. Finally, concluding remarks are provided in section VIII.

\section{Energy management in microgrids}

Effective management of energy in microgrids is significant for maintaining the stability, reliability, and efficiency of power systems. Different methods have been used in the literature to achieve proper energy management such as peak shaving, demand response, energy cost reduction and power loss minimization. This section sheds light on the conventional methods deployed for energy management, which include rule-based methods, optimization methods, optimized rule-based methods, before discussing the role of AI in improving these methods.

\subsection{Energy management applications}

 This section discusses important applications of energy management, like peak shaving, demand response, energy cost reduction, and power loss minimization.

\subsubsection{Peak shaving}
A major challenge for EMS in microgrids is keeping a balance between power generation and the load's power consumption. This becomes especially critical during peak demand hours. During these times, electrical lines and transformers may become strained, increasing the risk of equipment failure~\cite{b23}. Additionally, fuel consumption from generators, including backup ones, rises, leading to reduced operational efficiency. To address this, peak shaving is used to balance demand and supply, ensuring generators operate cost-effectively and with a higher load factor. By lowering peak demand, microgrids become more efficient and reliable, while electricity costs are also reduced. Peak shaving benefits both grid operators and consumers by improving the efficiency and reliability of both the microgrid and the utility grid. Various methods have been proposed in the literature to manage peak demand and maintain the balance during high-demand periods. These include energy storage systems~\cite{b24}, load management~\cite{b25}, on-site power generation using renewable energy sources (RES) or backup generators, and demand response programs that incentivize customers to reduce consumption during peak hours.

One way of obtaining peak shaving is the use of battery energy storage systems (BESS)~\cite{b26}. The process of storing excess energy generated by RESs and delivering whenever required is done by BESS. BESS help in the increased use of solar energy, thereby, increasing the self-consumption of grid-connected microgrids. The charging and discharging schedules of BESS, when used for peak shaving, are managed by different methods, for example, dynamic programming, rule-based algorithms, and genetic-based algorithms, etc.~\cite{b27,b28}. Another method of achieving peak shaving is by adding more equipment like transformers, cables, or distribution panels to the existing infrastructure to manage higher peak loads. This method does not directly reduce peak demand, but it makes the system more stable and reliable to manage the peak demand without any overloading issues. 
Peak shaving is also achieved through the use of on-site power generation systems like solar panels, gas or diesel generators, or CHP systems to decrease the load demand on the grid during peak times by producing its own electricity to meet the needs. However, this method may not be cost-effective and may lead to increased carbon emissions, high fuel consumption, extra maintenance costs, and equipment degradation~\cite{b29}-\cite{b31}. Thus, peak load shaving emerges as a feasible alternative where consumers can shift their peak consumption from a high-priced time to a low-priced time-period to save their electricity costs.  

\subsubsection{Demand response}
Demand response (DR) plays a very significant role in energy management by shaving the electricity demand peaks and ensuring smooth energy flow within the grid. accordingly, it is expected that DR is set to become a part of the energy system~\cite{b32}. Adjusting the electricity demand in response to the generated supply constitutes DR, which can be categorized into two main types: incentive-based and price-based. 

In the incentive based DR, the consumers get incentives for reducing the energy consumption during the peak time. One type of incentive-based DR is the direct load control (DLC) program, where small-scale consumers, like in residential areas, give control to the utility to control specific loads utilized by the end-users.  In these type of DR programs, a contract is made where the duration and number of interruptions will be defined so as not to create a major disturbance to the consumer. In return, the consumers receive compensation on their electricity bills and even other incentives like discounts~\cite{b33}. However, in the DLC program, the contract is prepared by the utility and the consumer is not notified of the interruption beforehand. Another type of DR is the curtailable load program, which targets medium to large consumers. In the curtailable load program, the consumers receive incentives to switch off certain loads and also by answering the utility calls to interrupt their energy supply. Unlike the DLC program, the curtailable load program is mandatory, and the consumers may have a penalty imposed if they fail to answer the utility when needed for DR~\cite{b33,b34}. 

In the price-based DR, time of use pricing is used where consumers are charged different rates of electricity supply cost in different time period within a day. The time of use pricing generally has an off-peak rate, peak rate and sometimes a shoulder-peak rate for different time period as per the utility~\cite{b33}. Time of-use pricing is a long-term electricity provision cost related to the use of electricity during a specific period of time in a day. However, to tackle the short-term costs associated with the electricity supply during critical times for the power system, critical peak pricing scheme is used. The maximum number of days per year and number of periods at which critical peak pricing will be applied will be defined by the utility.

\subsubsection{Energy cost reduction}

In microgrids, energy cost reduction is very significant to achieve economic profitability. Energy cost reduction is achieved by designing and employing strategies that help in reducing the cost of energy consumption and generation. By employing optimization techniques in microgrids for energy generation, storage and consumption based on real-time pricing and load forecasts, economic feasibility can be attained. Optimization strategies for DC microgrids, emphasizing cost reduction under real-time pricing conditions are discussed in~\cite{b35}. Another work done in~\cite{b36} presents the methods to minimize operating costs in DC microgrids through adaptive algorithms under real-time pricing scenarios. Research in~\cite{b37} focuses on economic dispatch strategies to reduce operating costs in microgrids under real-time pricing and in~\cite{b38}, a multi-objective optimization scheduling model is proposed for microgrids, integrating real-time meteorological data and load forecasting to achieve comprehensive energy management. Advanced metering infrastructure, data analytics in real-time and automation can help improve in dynamic pricing and forecasting load conditions of the microgrid. Automated demand response systems can optimize the energy consumption to avoid high-costs and the use of advanced metering infrastructure.

 Costs can be reduced significantly by utilizing RES and by reducing the microgrid’s dependency on the main grid during the peak hours. A novel approach that incorporates demand response to reduce operating costs in microgrids is presented in~\cite{b39} that is equipped with RES and ESS. Research in~\cite{b40} introduces a stochastic optimization model that accounts for the variability of wind and photovoltaic power within a microgrid. In~\cite{b41}, a two-layer optimal dispatch model for microgrids is developed, taking into account the uncertainties associated with wind power, photovoltaic generation, and load demand. ESS like lithium-ion batteries also play a key role in reducing costs. The storage systems can be charged during low-price period and discharged during high-priced period. Battery management systems also come useful in reducing costs by maximizing the efficiency of the battery. A dynamic economic dispatch model for microgrids that incorporates battery energy storage systems (BESS) is proposed in~\cite{b42}. The model addresses the time-coupling characteristics of BESS and the uncertainties of renewable energy sources. In~\cite{b43}, a formulation to determine the appropriate power dispatch of an energy storage system, considering the dependency on previous charging/discharging patterns. The algorithm is applicable in both grid-connected and islanded modes of operation of microgrid. A two-layer predictive energy management system for microgrids is introduced in~\cite{b44}, equipped with hybrid energy storage systems, considering degradation costs to optimize economic operation.  

\subsubsection{Power loss minimization}

Designing controllers for microgrids leads to concerns with the power balance, system stability and optimization as compared to the conventional grid. Specifically with respect to system optimality, various problems arise such as power loss in the transmission lines, battery life expectancy, power generation cost, greenhouse gas emission, etc.~\cite{b45}. Moreover, as microgrids incorporate different types of DERs, which leads to problems like reverse power flow, increase in line losses, over-voltage, etc.~\cite{b46}. Therefore, it is essential to reduce the power loss to improve the reliability and sustainability of microgrids. Effective use of power loss minimization methods can help improve energy routing, reduce operational costs, and improve the overall performance of the microgrid.

In~\cite{b47}, an optimal allocation of distributed generators is done in the microgrid for power loss reduction and voltage stability improvement by proposing the water wave optimization (WWO) algorithm. An adaptive active power optimization strategy is proposed in~\cite{b48} to minimize power losses during power transmission in islanded microgrids. The use of efficient energy routing algorithms can help in the reduction of power loss ~\cite{b49}. A hierarchical power routing scheme for Interlinking Converters (ICs) in unbalanced hybrid AC-DC microgrids is proposed in~\cite{b49}. This approach focuses on minimizing power losses within the microgrid while also improving system loadability at the point of common coupling (PCC). Study in~\cite{b50} focuses on developing an energy routing scheme for path selection in smart grids and the energy internet. A genetic algorithm is employed to calculate power losses and identify the optimal energy transfer route, considering losses occurring during both transmission and conversion processes. Reducing line resistance can also reduce power-loss in microgrids. Modeling of the distribution power loss in islanded three-phase AC microgrids is done in~\cite{b51} by incorporating both line losses and power converter losses, as a quadratic function of current allocation coefficients. This approach enables real-time efficiency optimization control of the AC microgrid, irrespective of variations in line resistance and load conditions.

\subsection{Energy management techniques} 

The main energy management techniques employed in microgrids can be categorized into rule-based, optimization-based, and hybrid approaches.

\subsubsection{Rule based methods}

Rule-based methods are simple techniques used for energy management in microgrids that are based on pre-defined conditions set by system designers and operators. These methods depend on a set of "if-else" instructions tied to initial conditions and rules. While easy to implement and require minimal computational effort, they cannot provide optimized solutions compared to other advanced methods~\cite{b52}. However, their simplicity limits their ability to handle complex scenarios.

An improved rule-based energy management system is proposed in~\cite{b53} for an off-grid microgrid comprising energy storage systems (ESS), wind turbines (WT), photovoltaic (PV) systems, and diesel generators. The results showed a reduction in operational costs by \$2827.96 and power losses by 1742.77kW. Similarly, in~\cite{b54}, a rule-based algorithm was used to implement an EMS that increases RES integration and manages power flow in the microgrid. This system was further optimized using the grasshopper optimization algorithm (GOA) for long-term capacity planning in an islanded microgrid. Another study in~\cite{b55} proposed an iterative rule-based EMS for a standalone microgrid consisting of PV panels, a diesel generator, lithium-ion battery storage, and a tidal turbine (TT). The study focused on finding the optimal sizes for PV-TT-battery components to minimize energy costs, greenhouse gas emissions, and power losses from electronic converters.

\subsubsection{Optimization methods}

Optimization methods often use mathematical models and algorithms and aim to find the best solution for a given problem.  In the context of RES design, optimization problems generally involve minimizing costs related to energy generation and supply, reducing power losses and optimizing the use of RES. Compared to rule-based methods, optimization-based methods can better adapt to changes in energy demand and supply. The operational costs of a grid-connected microgrid were minimized in~\cite{b56} by using the Harris Hawks optimization method, effectively coordinated energy management among distributed generation sources, enhancing economic efficiency. In another work done in~\cite{b57}, the economic load dispatch was done by using particle swarm optimization, which reduced the total costs by 17.2\% and total losses by 94.87\% as compared to classical methods. Optimization of the EMS of an on-grid microgrid was done in~\cite{b58} using the branch and bound technique. By using this technique, the optimal value of the state of charge (SoC) during the daily system operation was found to reduce the cash flow.
Heymann et al. ~\cite{b59} proposed an optimal EMS for standalone microgrids using Bellman's dynamic programming method. Their EMS model incorporated the operational costs of conventional generators and load shedding within its objective function. This study compared the performance and efficiency of the proposed method against classical nonlinear programming and mixed-integer linear programming models, evaluating both operational costs and computational time. The results showed that the proposed dynamic programming method was more effective than the traditional methods.

\subsubsection{Optimized rule-based methods}

Optimized rule-based energy management methods in microgrids combine the optimization techniques with predefined rules to improve efficiency, stability, and cost-effectiveness. These methods are specifically effective in the integration of renewable energy sources and energy storage systems within microgrids. A study by Chakraborty et.al~\cite{b60} introduced a cost-optimized, reliable, resilient, real-time rule-based energy management scheme for a solar photovoltaic array and battery energy storage (BES) integrated, grid-connected microgrid. This method aimed to address economic challenges by providing power leveling, emergency backup, and optimized power consumption from the grid. A multi-objective EMS for a remote microgrid is introduced in~\cite{b61}, focusing on economic load dispatch and battery degradation costs. This system makes use of genetic algorithms for day-ahead planning and a rule-based method for real-time operations. During real-time management, it consecutively considers diesel generator output, battery usage, and load shedding to maintain the balance between load and generation.

The analysis of MG EMSs based on methods such as rule-based, optimization and optimized rule-based methods is summarized in Table 1. In summary, traditional energy management techniques have their strengths and weaknesses. Rule-based methods are simple and easy to implement but lack flexibility. Optimization methods offer better adaptability and performance but require more resources and data accuracy. Optimized rule-based methods are balanced by incorporating optimization into rule-based systems to enhance efficiency and adaptability.
\sloppy

\begin{longtable}{|P{1.5cm}|P{4.2cm}|P{4cm}|P{4cm}|}
  \caption{Analysis of MG EMSs based on different energy management techniques} \\
  \hline
  \textbf{Ref.} & \textbf{Proposed EMS approach} & \textbf{Prospects} & \textbf{Limitations} \\
  \hline
  \endfirsthead
  
  \hline
  \textbf{Ref.} & \textbf{Proposed EMS approach} & \textbf{Prospects} & \textbf{Limitations} \\
  \hline
  \endhead

 \cite{b53} & Rule-based method & Reduced operational costs and minimized power losses. & Scalability and DR are not considered.\\
\hline
\cite{b54} & Rule-based method & Increase the integration of RES and manage the power flow of the proposed microgrid components. & The model lacks the consideration
of intermittent nature of RES and load demand fluctuations.
\\
\hline
\cite{b55} & Rule-based method & Optimal PV-TT-battery sizing is obtained to achieve a minimum cost of energy, cost of greenhouse gas emission, and cost of power losses caused by power electronic converters. & Scalability and load demand is not considered.
\\
\hline
\cite{b62} & Rule-based method & A central EMS oversees the overall microgrid's operations, while the prosumer EMS addresses power imbalances within individual prosumer systems. & Demand response approaches and green house gas emissions are not considered.
\\
\hline
\cite{b47} & Optimization-based method  & Optimal allocation of distributed generators is done in the microgrid for power loss reduction and voltage stability improvement. & Economic factors and intermittent nature of RES are not considered.
\\
\hline
\cite{b56} & Optimization-based method  & Effectively coordinates energy management among distributed generation sources, enhancing economic efficiency. & Scalability and economic factors are not considered. \\
\hline
\cite{b57} & Optimization based method  & The economic load dispatch was which reduced the total costs and total losses. & Real-time implementation and RES’ intermittent nature is not considered. \\
\hline
 \cite{b58} & Optimization-based method  & Optimal value of the state of charge (SoC) during the daily system operation is found to reduce the cash flow. & Computational complexity and scalability are not considered. \\
\hline
 \cite{b59} & Optimization-based method  & Operational costs and computational time is reduced. & Computational complexity and scalability are not considered. Simplified component modeling.  \\
\hline
\cite{b60} & Optimized rule-based method  & cost-optimized, reliable, real-time energy management scheme for grid-connected microgrid. & Intermittent nature of RES is not considered and limited adaptability to load demand.  \\
\hline
\cite{b61} & Optimized rule-based method & Economic load dispatch and reduced battery degradation costs. & Scalability and simplified battery degradation model. \\
\hline

\end{longtable}
\fussy
\section{Overview of Artificial Intelligence (AI)}
 AI has the potential to tackle the complexity associated with handling and processing massive amounts of data and automating the decision-making process in power systems due to its ability to execute sophisticated analytics on data, identify patterns to predict maintenance needs and optimize the distribution of energy. The potential benefits for microgrids include an increase in efficiency, reliability, and robustness introduced by utilizing the benefits of AI’s ability to learn from big data and make real-time decisions~\cite{b63}. These prospects are especially beneficial for energy storage and supply management, for making use of RERs and for the optimization of the operating expenses to make the energy structure more efficient and eco-friendly. Artificial intelligence tools have shown strong merits in providing eminent solutions for tackling such challenges and optimizing the performance of distributed energy systems.

\begin{figure}
    \centering
    \includegraphics[width=1\linewidth]{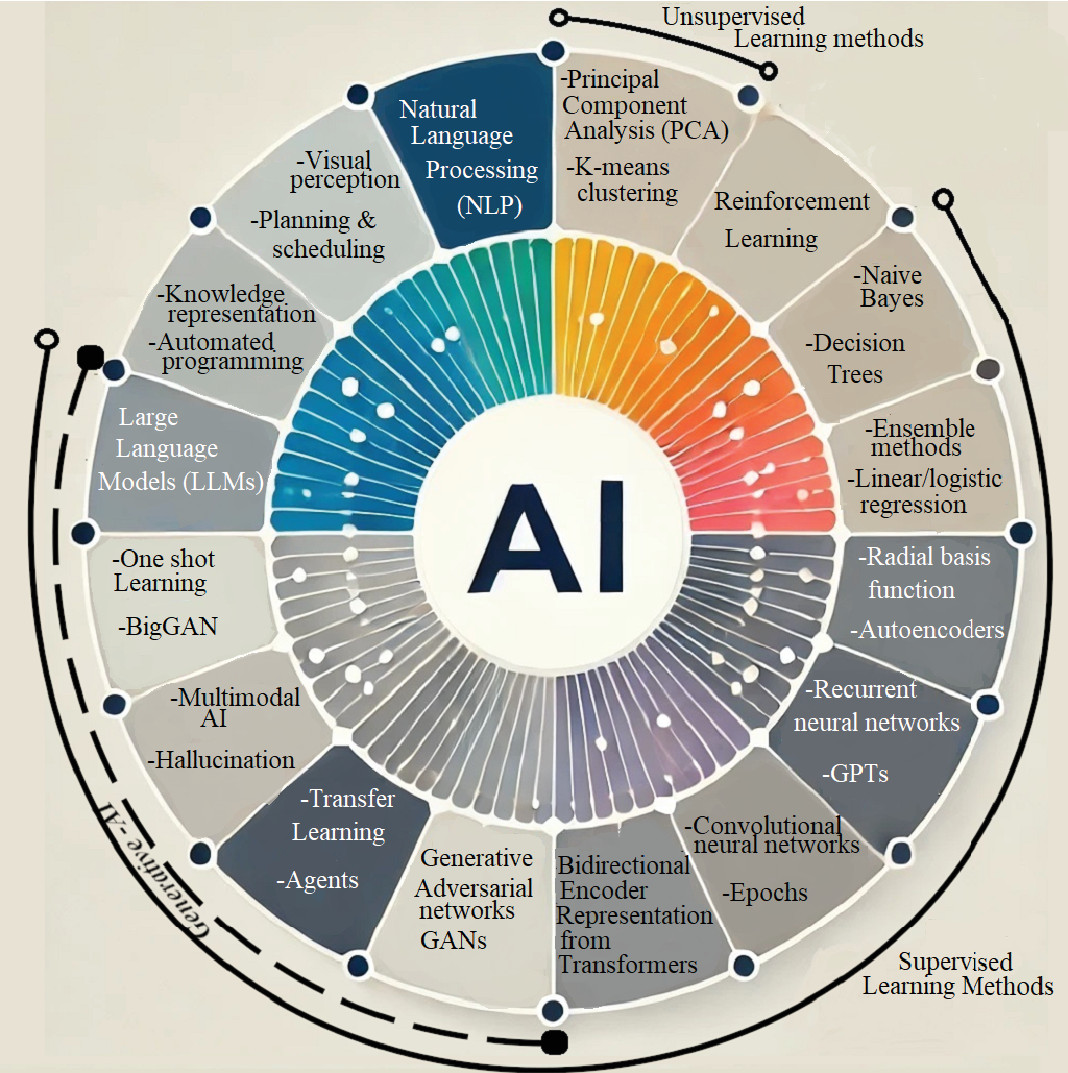}
    \caption{Basic wheel diagram of AI and machine learning methods}
    \label{fig: Figure 2}
\end{figure}

Artificial intelligence, and in particular machine learning, contributes towards the creation of algorithms and statistical models with the help of which the machine can learn and make a decision or make a prediction based on it~\cite{b64}. A wheel diagram of AI is depicted in Fig. \ref{fig: Figure 2}. Unlike conventional programming that requires an individual to instruct the machine what to do and how to do it in the best way, machine learning lets the system learn on its own from the available data and enhance its proficiency in handling future data without the need to program it for every scenario. Machine Learning (ML) algorithms consist of three different types~\cite{b65}. The first type is unsupervised learning, where the training of the algorithms is done using unlabelled datasets to find trends and details, while the second type is supervised learning, in which labeled datasets are used to train the algorithms to predict the output~\cite{b66}. The third approach is reinforcement learning (RL), in which an agent uses trial and error to learn from its environment, aiming to maximize its rewards over time~\cite{b67}.

Deep learning is a branch of machine learning that focuses on algorithms that imitate the structure and functioning of the human brain's neural network. It uses stacked artificial neural networks (ANNs) to identify complex patterns in large datasets~\cite{b68}. These networks consist of multiple layers: an input layer, hidden layers, and an output layer. The hidden layers process data through weighted connections, and functions like rectified linear unit (ReLU) and sigmoid introduce non-linearity, enabling the model to learn complex features. Forward propagation computes the output, the loss function measures errors, and back-propagation adjusts weights and reduces errors using optimization techniques like gradient descent~\cite{b68}. Together, these components allow deep learning models to analyze data and make predictions.

Ensemble methods is another category that involves techniques that combine the outcomes of multiple AI algorithms~\cite{b69}. This approach is used to address the limitations of individual algorithms, leading to improved overall results. Expert Systems (ES), as an established branch of AI, are computer-based models imitating the decision-making skills of humans, which enables them to solve problems effectively~\cite{b70}. These systems can be used in many power system applications to increase their efficiency, performance, and other important features.

\subsection{Deep learning architectures}
Deep learning (DL) and artificial neural networks (ANNs) are closely related concepts that have advanced greatly in the field of AI. DL is a subset of machine learning which focuses on training deep neural networks (DNNs) with multiple layers to learn hierarchical data representations. ANNs form the foundational structure for DL algorithms which makes the development of complex models capable of extracting complex patterns and making precise predictions from large datasets. Different DL architectures have been developed over time to solve problems in different areas. Among the common architectures of deep neural networks, we cite Convolutional Neural Networks (CNN), Autoencoders, Restricted Boltzmann Machines (RBM) and Long Short-Term Memory (LSTM)~\cite{b68}.

\subsubsection{Convolutional neural networks}
Convolutional Neural Networks (CNNs) are inspired by the human visual cortex and are widely used in image analysis within computer vision. A typical CNN includes layers of convolution and sub-sampling, followed by a fully connected layer and a softmax function layer. Features are extracted at each convolutional layer from the input to the output. Between convolutional layers, pooling or sub-sampling layers are added. The fully connected layers are responsible for classification. The input to a CNN is a 2D image with N-by-N pixels. Each layer contains groups of 2D neurons, referred to as kernels or filters. In CNNs, neurons in each convolutional layer do not connect to all neurons in the next feature extraction layer. Instead, they connect only to partially overlapping neurons in the feature map of the previous layer. This overlapping area is called the local receptive field. All neurons in convolutional layers share the same number of connections to the previous layer and have identical weights and biases, making learning faster and memory requirements lower. Each neuron looks for similar patterns in different parts of the input image. Sub-sampling layers reduce the network's size using local averaging filters and max or mean pooling~\cite{b71}.

In the final layers, neurons are fully connected and handle classification. A deep CNN is created by stacking multiple convolutional and sub-pooling layers with shared weights. This deep structure produces high-quality representations while minimizing the number of parameters~\cite{b72}.The algorithm consists of a cost function which measures the difference between the predicted output $h_{w,b}(x)$ and the actual output $y$ for an individual training sample $(x,y)$ in hidden layers, which is defined as~\cite{b72}: 
\begin{equation}
J (W, b; x, y) = \frac{1}{2}||h_{w,b}(x)-y| |^2
\end{equation}
The error term $\delta$ for layer $l$ is given by the equation [31]:
\begin{equation}
\delta^{(l)} = \left( {W}^{(l)} \right)^{\mathrm{T}} \delta^{(l+1)} \cdot f'\left(z^{(l)} \right)
\end{equation}
where $W$ represents the matrices containing the weights for each layer in the neural network. Each $W^{(l)}$ connects layer $l-l$ to layer $l$ and $b$ represents vectors containing the biases for each layer. Each $b^{(l)}$ is added to the weighted inputs of layer $l$. $\delta^{(l+1)}$ is the error for $(l + 1)$th layer of a network whose cost function, $J (W, b; x, y). f' (z ^{(l)} )$ represents the derivate of the activation function.
\begin{equation}
\nabla _{w^{(l)}} J(W, b; x, y) = \delta^{(l+1)} \left( a^{(l+1)} \right)^T
\end{equation}
\begin{equation}
\nabla _{b^{(l)}} J(W, b; x, y) = \delta^{(l+1)}
\end{equation}
where $a$ is the input, such that $a^{(1)}$ is the input for the 1st layer
which is the actual input image and $a^{(l)}$ is the input for $l$th
layer. The gradients indicate how much the weights and biases should be adjusted to minimize the cost function. These equations form the core of the backpropagation algorithm, which trains neural networks by optimizing the cost function via gradient descent.

\subsubsection{Autoencoders}

Autoencoders are neural networks that follow the unsupervised learning approach. Autoencoders learn the representation from the input data set to reproduce the original data set with dimensionality reduction.  The learning algorithm is based on the idea of backpropagation. Autoencoders follow the basic principles of principal component analysis (PCA). A PCA converts multi-dimensional data into a linear representation, whereas autoencoders do a bit more by converting it into a non-linear representation. PCA finds a set of linear variables in the directions with largest variance. PCA finds the variances instead of covariance and correlations, as it looks for linear representations with the most variance~\cite{b73}. The main aim is to look for the direction with the least mean square error, which has the least reconstruction error. Autoencoders use encoder and decoder components with nonlinear hidden layers to extend the capabilities of PCA. This design helps in dimensionality reduction and to reconstruct the original data. Autoencoders are considered unsupervised deep neural networks, although autoencoders use backpropagation, which is generally used for supervised learning. This classification is done because the goal of autoencoders is to reconstruct their input $x^{(i)}$ rather than just predicting distinct target values $y^{(i)}$; in essence, $y^{(i)} = x^{(i)}$. Also, Hinton and colleagues~\cite{b74} demonstrated that autoencoders could achieve near-perfect reconstruction of 784-pixel images, outperforming PCA.

\begin{figure}
    \centering
    \includegraphics[width=1\linewidth]{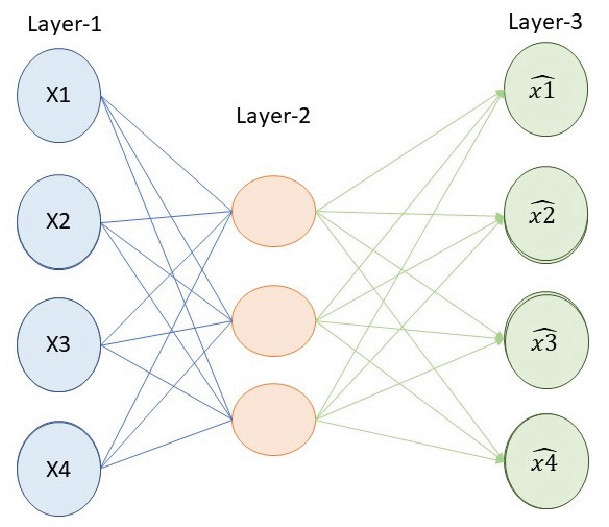}
    \caption{Representation of nodes of autoencoders}
    \label{fig: Figure 3}
\end{figure}

Figure \ref{fig: Figure 3} shows a simplified autoencoder architecture that reduces input data dimensionality and reconstructs it at the output layer. A deep autoencoder was implemented in~\cite{b75} with stacked restricted Boltzmann machine (RBM) blocks to achieve better modeling accuracy and efficiency compared to the proper orthogonal decomposition method for dimensionality reduction in distributed parameter systems.

\subsubsection{Restricted Boltzmann Machine}
RBM is an artificial neural network used to produce non-linear generative models by using unlabeled data. The aim is to train the network to maximize a function of the probability distribution in the visible units which will help in reconstructing the input data using probability. RBM comprises of two layers: hidden layer and visible layer. In the visible layer, every unit is connected to all units in the hidden layer but the units in the same layer are not connected. The energy (E) function of the configuration of the visible and hidden units, $(v, h)$ is expressed as below~\cite{b74}:
\begin{equation}
E(v, h) = - \sum_{i \in \text{visible}} a_i v_i - \sum_{j \in \text{hidden}} b_j h_j - \sum_{i,j} v_i h_j w_{ij}
\end{equation}
where, $v_i$ and $h_j$ are the vector states of the visible unit, $i$ and hidden
unit, $j$. $a_i$  and $b_j$ represent the bias of visible and hidden units.
$W_{ij}$ denotes the weight between the respective visible and hidden units.

 \subsubsection{Long Short-Term Memory}

LSTM was first proposed by Hochreiter et al.~\cite{b76} in 1997 and is specialized architecture of recurrent neural networks (RNNs) designed to effectively capture and learn long-term dependencies in sequential data. LSTM can store values from previous states and can be trained for tasks that require memory or to hold the current states. LSTM networks address an important limitation of RNNs which is the vanishing gradient problem~\cite{b77}. LSTMs achieve this by allowing gradients to pass through unaltered, effectively preserving long-term dependencies during training. An LSTM comprises of memory cell states through which the signals flow, regulated by input, forget, and output gates. These gates control the information stored, read, and written within the cell. Notably, many major companies use LSTM architectures in their voice recognition platforms~\cite{b77}. The computations of the gates are described in the equations below~\cite{b76,b78}:
\begin{equation}
f_t = \sigma(W_f \cdot x_t + w_f \cdot h_{t-1} + b_f)
\end{equation}

\begin{equation}
i_t = \sigma(W_i x_t + w_i h_{t-1} + b_i)
\end{equation}

\begin{equation}
o_t = \sigma(W_o x_t + w_o h_{t-1} + b_o)
\end{equation}

\begin{equation}
c_t = f_t \odot c_{t-1} + i_t \odot \sigma_c(W_c x_t + w_c h_{t-1} + b_c)
\end{equation}

\begin{equation}
h_t = o_t \odot \sigma_h(c_t)
\end{equation} 
where $f$,$i$, and $o$ are the forget, input, and output gate vectors, respectively. $W$, $w$, $b$, and $\odot$  represent weights of input, weights of recurrent output, bias, and element-wise multiplication, respectively.

There is another smaller variation of the LSTM known as gated recurrent units (GRU). GRUs are smaller than LSTM in size as they do not have the output gate and can perform better than LSTMs on some simpler datasets~\cite{b79,b80}. LSTMs recurrent neural networks can keep track of long-term dependencies.
LSTMs are suited for building models that depend on context and on previous states. The cell block of LSTMs stores information of previous states. Decisions regarding the new data going into the cell, what remains in the cell and the cell values used in the obtaining the value of the output of the LSTM block are controlled by the input, forget and output gates respectively~\cite{b76,b78}. Naul et al. demonstrated LSTM and GRU based autoencoders for automatic feature extractions~\cite{b81}.

\subsubsection{Generative adversarial networks}
Another powerful deep learning technique called generative adversarial networks (GANs) aim at producing new and high-quality data that are very similar to what has been presented during training. GANs' operation is based on two neural network models: the generative model, also known as the generator, and the adversarial model, or the discriminative model, also known as the discriminator~\cite{b82}. The generator generates new data from noise and tries to generate results that cannot be differentiated from actual data. At the same time, the discriminator decides between real and synthetic data they were trained on, trying to determine which is fake. Training involves the generator getting smarter in its data generation process to make the discriminator believe its generated data is real, while the discriminator gets smarter in its ability to determine real data from the fake data generated by the generator. This min-max optimization is carried out until the generator has generated a near-realistic synthetic data set, which makes it hard for the discriminator to differentiate.

Table 2 shows a comparison between different deep-learning networks. A variety of algorithms based on AI in microgrids have, in fact, been investigated in the literature for a range of energy management tasks, as illustrated next.
\sloppy
\begin{table*} 
\centering
   \caption{ Comparison of different deep-learning networks}

\label{tab2}
\begin{tabular} {|P{1 cm}|P{3 cm}|P{3 cm}|P{3 cm}|P{4 cm}|}
\hline
\textbf{{Ref.}} & \textbf{{Architecture}} & \textbf{{Network model }} & \textbf{{Type of Training}} & \textbf{{Training Algorithm}} \\

\hline

 ~\cite{b71}& CNN & Discriminative	&Supervised Training &	Gradient descent based backpropagation   \\
\hline
 ~\cite{b73}&Autoencoder & 	Generative	& Unsupervised Training	& Backpropagation\\
\hline
 ~\cite{b74}&RBM	&Generative with discriminative fine-tuning	&Unsupervised Training&	Gradient descent \\
\hline
 ~\cite{b82}&Adversarial Networks&	Generative and discriminative&	Unsupervised Training&	Backpropagation\\
\hline
 ~\cite{b79}&LSTM&	Discriminative&	Supervised Training& 	Gradient descent and backpropagation through time \\
\hline
  ~\cite{b87}&RBF&	Discriminative&	Both Supervised and Unsupervised Trainings &	Least square function, K-means clustering  \\

\hline

\end{tabular}

\end{table*}
\fussy
\section{AI enabled microgrid prospects}
Merging AI techniques into EMS of microgrids guarantees a promising approach towards increasing the efficiency, reliability, and sustainability of these distributed energy systems~\cite{b18}. This section discusses the powerful capabilities of AI to effectively address the main challenges associated with microgrids, particularly the uncertain and unpredictable nature of RES; the need for real-time optimization of energy consumption and production, and protection from cyber-attacks. 

\subsection{Enhancing energy efficiency, demand management and reduced operational costs}

AI can optimize energy utilization within microgrids (MGs) by opportunistically balancing demand and supply in real-time~\cite{b83}. AI-powered EMS may consider factors such as consumer behavior, energy prices, and grid conditions to make better decisions about energy dispatch, storage, and demand response. This optimization capability that easily evolves to changes can lead to major reductions in energy costs and improvements in grid stability.

The data-driven control techniques are considered as essential tools in improving energy management of the  MGs. Machine-learning methods (among other artificial intelligence tools) can be used since these methods are beneficial to process a massive amount of data in forecasting renewable energy outputs, managing demand-side load response to subsequently optimize the generation and storage, and distribution of energy. The model-free EMS approach followed in~\cite{b84} has already shown good performance in different energy management tasks making MGs more efficient, reliable, and resilient. Incorporating renewable energy sources within MGs presents a significant challenge in controlling their uncertain and unpredictable output.

To address the challenge of controlling the uncertain nature of renewable energy resources (RERs), AI tool such as generative adversarial networks, have been used in~\cite{b84}. With the help of GANs, a model for unpredictable output from RERs was developed and subsequently, a data-driven EMS was designed for the effective management of energy resources in a microgrid. In this regard, the EMS considers the maintenance costs associated with the operation of DERs, besides energy management, and their reactive power capabilities to make economically viable and smooth decisions in the energy management process. GANs have similarly shown to be effective in carrying out optimum energy management in the operation of microgrids through modeling the behavior of consumers and the dynamic behavior of the RES. The elaborate insights into energy demand and supply patterns from this GANS-generated information helps optimizing the operation of MGs.

Another approach in~\cite{b85} deals with an EMS developed for optimum energy management in an islanded microgrid. The system integrates a multiperiod artificial bee colony (MABC) approach with Markov chain modeling to yield a combination of efficiency and reliability in energy delivery, where microgrids work independently of the primary grid. The MABC algorithm approach to the optimization technique works based on honeybees' searching behavior to find the optimum in multiple periods. It is well-suited to solve complex optimizations with dynamic and uncertain conditions. A Markov chain is used to facilitate the random behavior that involves energy demand and supply in the microgrid. It foresees the future state based on the current condition, thus helping the system by providing probabilistic analysis. The joint MABC and Markov chain approach in~\cite{b85} leads to better dynamic performance of EMS in an islanded microgrid. This system ensures that the various available energy resources are effectively managed for optimized performance, minimization of the operating cost, and constant availability of power supply when it is needed. Experimental and simulation results show a reduction in total operation costs by about 30\% and a lesser market clearing price in each time interval due to appropriate and real-time control of demand response through the proposed algorithm. 

In~\cite{b86}, a model is introduced to optimize the power supply, minimize the intake from the utility grid and maximize the supply from RERs. Research carried out by Urias et al.~\cite{b86} employs a Recurrent Neural Network (RNN) architecture as a key component of the solution. RNNs are especially useful in learning from and processing sequential data. This makes RNNs well-suited for modeling the dynamic nature of energy systems. The authors’ approach in~\cite{b86} treats each energy source as an agent, and a renewable multi-agent system is created by combining them. In order to achieve the optimal operation of a microgrid
interconnected to a utility grid, an optimization problem is modeled to determine the optimal power values for a
time horizon of one week for wind, solar, and battery systems
and the utility grid. The optimization technique takes into account multiple factors, including the generation of solar and wind power, the power requirements from the utility, the limitations imposed by the utility grid, and the charging status of batteries. The neural network is trained with time series
data as an input using an extended Kalman filter. This training approach enables the network to learn from historical data and make predictions regarding future energy demand. Simulations are then conducted to determine the power output of wind, photovoltaic, and battery systems over various one-week time-frames. The results of these simulations show high accuracy of the method proposed in optimizing the performance of renewable energy systems.

Reinforcement learning (RL) is used to solve the problem of distributed economic dispatch by working with sequential data in~\cite{b87}. ED is the process of allocating electrical power generation to different power plants in a way that minimizes costs and satisfies demand constraints. RL is a machine learning algorithm that uses trial and error to allow agents to learn how to behave in an environment. The authors of~\cite{b87} use a hierarchical RL (HRL) technique to handle the ED problem. In HRL, a complex problem is decomposed into smaller subproblems. The method of approximation used in~\cite{b87} is Radial Basis Function (RBF). RBF approximation involves the utilization of a set of basis functions to approximate functions. Reference~\cite{b87} experimentally demonstrates that the proposed HRL algorithm decreases computational cost and enhances learning efficiency as compared to traditional RL algorithms. In order to verify that their proposed hierarchical learning can decrease the costs of operations in the long run and enhance the stability of operations, simulations were conducted.

Another emerging energy management technique in MGs is Graph Convolutional Networks (GCNs). GCNs are used to develop probabilistic power flow methods that are computationally effective and accurate~\cite{b88}. Using GCNs to develop probabilistic power flow methods contradicts the traditional Monte-Carlo method, which may be computationally expensive. Power flow calculations are the foundations of any power system. GCNs were used in~\cite{b88} for the probabilistic power flow problem to increase the computational accuracy. In the proposed GCN framework,  only the power grid topology is sufficient, and no electrical knowledge is needed as an input. It significantly reduces the time of computing while its accuracy is high compared to those of traditional Monte Carlo simulation methods or other enhanced methods. This proposed approach can be widely applied in power system planning, operation, and dispatching.

\subsection{Improving forecasting }
Given the capabilities of deep learning (DL) in automating the design process by means of acting as a universal function approximator~\cite{b89}, AI can precisely predict RES output and energy demand, which enables EMS to adjust grid operations and optimize energy storage. The past data and weather patterns to forecast RES output can be analyzed using machine learning algorithms for higher efficiency, allowing EMS to plan energy storage and demand schedules accordingly. The sensor data analysis using AI also facilitates up-to-date maintenance and prediction of potential equipment failures and ensures smoother operations with lesser downtime.

Feedforward neural networks (NNs) belong to a specific category of neural networks characterized by the one-way flow of information, where data is processed and transmitted in a singular direction~\cite{b90}. In contrast to RNNs, feedforward neural networks lack cyclic structures among nodes. The research outlined in~\cite{b91} uses the capabilities of a feedforward neural network mechanism to tackle the task of feature selection. This effort of tackling the feature selection task aims to identify and extract significant predictors essential for improving the accuracy of forecasting electricity demand which is short-term within distributed energy systems, all achieved prior to the training phase.

The predictors or features, extracted using the proposed framework, are then employed as inputs for feedforward artificial neural networks, allowing for a comprehensive evaluation of forecast performance. The framework is designed and it integrates an algorithm, specifically genetic algorithm, for the precise process of feature selection~\cite{b91}. Furthermore, Gaussian process regression is utilized to assess the fitness scores of the selected features, contributing to a more refined understanding of their significance.

To experimentally verify this method, simulations were conducted over four different datasets, each observed to have a set of 24 features. The four electricity demand datasets are obtained from four types of buildings (customer classes) in the Otaniemi area of Espoo, Finland. The buildings are Building A (residential), Building B (educational, containing classrooms and laboratories), Building C (office), and Building D (mixed-use, containing computer laboratories and a healthcare center). The first set of features that are used in the model are calculated based on the characteristics of the traits of electricity usage in the decentralized energy systems, such as buildings, in conjunction with their historical data of consumption and the influence of external agents. These external agents include the market price of electricity, a minute/hour, month/season, weather conditions, and the presence or absence of people in the building. The model proposed reflects complexity in the sense that it can decide on the required or optimal number of features opting for a subset most strategically smaller than all from the actual number in the latent feature space. This reduction of features significantly improves the efficiency and effectiveness of the model. The proposed architecture in ~\cite{b91} is used to handle the associated challenges of forecasting short-term electricity demand in complex distributed energy systems. The improvement in forecast accuracy using binary genetic algorithm- Gaussian process regression feature selection (BGA-GPR FS) selected features to develop forecast model training inputs is 38.7\%, 81.2\%, 81.9\%, and 83.0\%, respectively. These experimental results further highlight the importance of effective feature selection in improving electricity demand forecasting. 

\subsection{Predictive maintenance}

Research carried out by Hatata et. al~\cite{b92} detected, classified, and located faults in the microgrid through Convolutional Neural Networks (CNN) optimization by the Gorilla Troops Optimization (GTO) technique. Using CNN, the processing is done into multidimensional arrays for current and voltage measurements in image classification, and the GTO fine-tunes the architecture and hyperparameters for its efficiency. The advanced adaptive protection (AP) scheme is introduced in ~\cite{b92} via implementing three different models: CNN-GTO-I for fault detection, CNN-GTO-II for fault type classification, and CNN-GTO-III for fault location. These models communicate over a dedicated channel, ensuring fast response to faults. Tests in a simulated environment of Future Renewable Electric Energy Delivery and Management (FREEDM) show high rates: 99.37\% for the detection model, 99\% for the classification model, and 98.2\% for the location model. This scheme also exhibits robustness against fault conditions and uncertainties compared to the conventional methods developed using SVM and fuzzy logic. Raspberry Pi and Internet of Things (IoT) platforms are used for practical implementation to demonstrate the scheme's effectiveness. 

Another study in~\cite{b93} explores maintenance strategies in islanded microgrids, focusing on both proactive and reactive approaches. By integrating fuzzy logic controllers and machine learning techniques, the self-healing capabilities of microgrids were improved, improving reliability and reducing downtime. A novel combination model of ANN and RBM was proposed in~\cite{b94} to detect and classify the faults for operations of both on-grid and off-grid microgrids. The proposed model fundamentally learned and classified unnatural signals to various faults faced by the microgrid.  A unique measurement of current and voltage waveforms using discrete wavelet transform against the changing patterns in-line parameters.  The results indicate that the proposed model achieved an accuracy of approximately 99.38\% in detecting and classifying short circuit faults for all types of faults.

\subsection{Enhancing microgrid resilience and cybersecurity}

Defensive strategies against data integrity attacks in smart grids are generally divided into two layers. The first layer includes techniques such as encryption, authentication, and authorization to prevent unauthorized access and ensure data integrity~\cite{b95,b96}. However, implementing these complex techniques directly on smart meters is a challenge due to their limited computational and storage capabilities. Therefore, the second layer of defense focuses on detection mechanisms deployed at control centers, which possess greater computational resources. These detection techniques are significant for identifying and reducing attacks that penetrate the initial defense layer. In the second layer, detection schemes play a significant role in defense against data integrity attacks, and many research efforts have been dedicated to developing these methods~\cite{b97}-\cite{b100}. Most of these techniques depend on data mining techniques~\cite{b101}.

Existing data mining detection methods are usually divided into three main types. The first type uses regression and prediction approaches~\cite{b99,b100}. These methods use past data to predict expected values. If the difference between the predicted value and the actual measured value is too large, the measured value is flagged as false (malicious). Otherwise, it is considered true (benign). For example, Ford et al.~\cite{b99} used Artificial Neural Networks (ANNs) to detect energy fraud. However, if the threshold for detecting differences is not set carefully, the accuracy of detection can be low.

The second type involves classification-based detection methods~\cite{b98,b102}. These methods learn rules from labeled historical data, which includes both normal and malicious examples. Once trained, the system can classify new data as true (benign) or false (malicious). For example, Faisal et al.~\cite{b97} created a system using stream mining to improve the security of Advanced Metering Infrastructure (AMI). However, these methods need labeled datasets, which often require expert knowledge. Getting such labeled data in real-world situations can be difficult, which may lower detection accuracy when expert input is not available.

The third type of detection methods is based on the Min-Max model~\cite{b103}. These methods determine a normal range for data by learning the minimum and maximum values from past records. If a new measurement falls outside this range, it is marked as false (malicious); otherwise, it is considered true (benign). For example, Baig et al.~\cite{b96} created a lightweight pattern-matching system to detect unusual device behavior in smart grids. However, if the data varies a lot, this method might set a very wide normal range, which can reduce the detection rate.

AI can improve the resilience of microgrids by enabling automated responses to disturbances and cyberattacks~\cite{b104}. AI-based fault detection and isolation systems can quickly identify and isolate faults, minimizing disruptions and preventing repeated failures. Additionally, AI can be employed to detect and mitigate cyberattacks targeting microgrid control systems, protecting the integrity and security of the grid.

GANs have also been applied to improve the security of microgrids. The work carried out by Tang et. al in~\cite{b105} has analyzed how hackers can change important data for microgrids that can lead to power outages or force people to reduce energy use. GANs can be used to find and stop these attacks, making microgrid EMS more reliable and safe. GANs are used to design the behavior of consumers and the varying nature of the output of renewable energy sources. The information gathered is then used to optimize the operation of the MG. The effect of cyber attacks on the data of the central control of MGs is checked in~\cite{b105}. Such attacks might lead to power outages, resulting in grid instability. This kind of unpredictable source modeling within an energy system and projecting human behavior toward the use of energy has set new ways in which to handle a large number of problems that occur while harnessing renewable sources of energy. Tang et. al~\cite{b105} explored using the power of GANs to make the EMS work even better for using energy more efficiently, involving stabilizing the grid and increasing the resilience of microgrids. As GANs continue to develop, they will play an even more significant role in the future of microgrid energy management. This may go an even longer way in helping design a more sustainable and resilient energy future.

Research carried out by Kang et. al in~\cite{b106} presents an innovative EMS incorporating ANNs aiming to regulate power within hybrid AC-DC distribution networks. This innovative EMS, driven by ANN technology, is specifically designed to gather crucial data, including information on distributed generation (DG), power supplied based on state of charge (SoC), and load demand. Accordingly, this system intelligently utilizes the collected data to make better decisions and select the most suitable way of operation. The proposed EMS harnesses the power of ANN which has been trained in a grid-connected mode, to exert control over each power converter, ensuring their optimal performance in their respective modes. This strategic application of ANN enables the EMS to dynamically adapt to diverse scenarios, thereby augmenting its overall efficiency and adaptability. To empirically validate the effectiveness of the proposed EMS, the hybrid AC/DC small-scale microgrid was utilized. Rigorous simulations and experiments were conducted across various operating modes, providing a comprehensive assessment of the system's performance and responsiveness. This experimental approach highlights the reliability and robustness of the ANN-based EMS in real-world scenarios, paving way for its potential application and integration into larger energy distribution networks.

The computational complexity of cybersecurity measures involves evaluating the time, processing power, and resources required to detect, prevent, and respond to cyber threats without overburdening the system. Advanced detection algorithms, such as sliding mode observers (SMOs), can effectively identify anomalies, but their complexity must allow real-time operation. For example,~\cite{b107} utilizes SMOs to protect DC microgrids from cyberattacks, specifically false data injection attacks (FDIAs). The approach used an advanced SMOs that accounts for unknown inputs to detect cyber threats in communication channels and measurement points. By improving the model of each potentially compromised DG unit with possible attack signals, a set of SMOs is employed for each DG to estimate the states of neighboring units. The real-time simulation of the direct current microgrid using the proposed method was executed in 0.8 microseconds on the OPAL-RT 5700 simulator. This demonstrates that the proposed approach effectively manages cyber-attacks in microgrids with numerous distributed generators without sacrificing performance. 
It is also important to balance the depth of security measures with available resources. Overly complex algorithms might strain the system, while simpler methods may fail to detect advanced threats. Scalability is another challenge as microgrids grow; cybersecurity protocols must adapt to larger, more complex systems.
Additionally, analyzing the potential impact of attacks through AI approach helps in developing robust countermeasures.~\cite{b108} addressed the critical challenge of maintaining precise voltage regulation in off-grid AC microgrids using an AI-based approach while enhancing their resilience against cyber-attacks, particularly FDIAs. The proposed method combines a secondary consensus control system with a deep neural network (DNN)-based observer to quickly identify and counteract FDIAs in microgrids. The consensus-based secondary control ensures accurate voltage regulation among DGs. The DNN-based observer proposed not only detects FDIAs with high precision but also reduces their impact, maintaining the microgrid's stability and reliability. Through extensive simulations and real-time testing using the OPAL-RT platform on an IEEE 34-bus test feeder system with varying numbers of DERs, the control strategy was shown to be both scalable and effective. 

Similarly, an artificial neural network (ANN)-based method to detect and address FDIAs in a DC microgrid was proposed in~\cite{b109}. The ANN is used to identify and calculate the value of the false data injected by attackers. This calculated value is then used to effectively eliminate the cyber-attack. The proposed method works quickly and performs efficiently to mitigate the attack. Unlike other approaches, this method does not require additional controllers or model-predictive controllers, which simplifies its implementation. The proposed technique was tested under various conditions, including load changes, communication delays, and the addition of new units (plug-and-play scenarios). Both constant and time varying FDIAs were considered during evaluation. Additionally, the ANN's performance was tested in a case study with white noise. The results show that this method can accurately calculate and remove false injected data rapidly and effectively. This detection and removal approach forms the foundation of a robust control strategy that ensures the system's stability and resilience against cyberattacks.

\section{Proof-of-concept: numerical evaluations}
\begin{figure}
    \centering
    \includegraphics[width=1\linewidth]{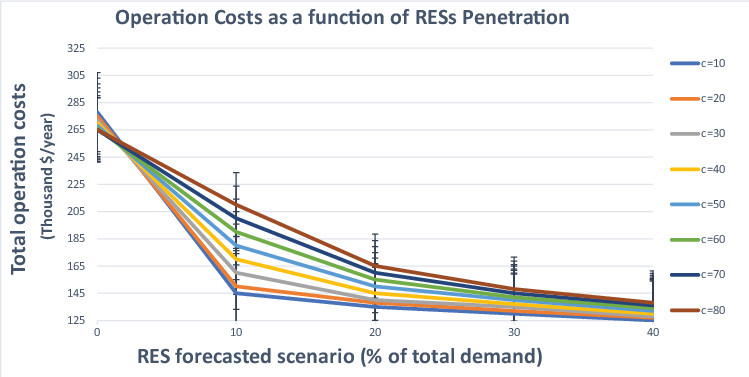}
    \caption{Operation costs as a function of penetration and uncertainty level of RESs as a percentage of the hourly total demand}
    \label{fig: Figure 4}
\end{figure}

This section presents numerical evaluations that illustrate
the potential of AI-based EMS in fulfilling the main microgrid applications presented in the previous section, i.e., in enhancing energy efficiency, demand management, and reducing operational costs, in improving forecasting and predictive maintenance, and in enhancing microgrid resilience and cybersecurity. The presented results are based on the use of GANs to model the uncertainties in the output power of the RESs and to further co-optimize the costs related to both active and reactive powers of diesel generators as proposed in~\cite{b84}. The uncertainties of the RESs are simplified into three cases: forecast, high and low. The power penetration level of the RESs, such as wind turbines or photovoltaics, is defined by setting the forecast case as a percentage of the hourly total demand (x-axis in Fig.~\ref{fig: Figure 4}). The low and high cases are then set up as percentages below and above the RESs power forecast, respectively. The level of uncertainty is structured in such a way that the low and high scenarios deviate from the forecast scenario. Uncertainty is smoothed out in increments of 10\%, so a $c$\% in uncertainty implies that the high and low scenarios are the RES's power forecast multiplied by ($1 +$ $\frac{c}{100}$) and ($1 -$ $\frac{c}{100}$), respectively. 

Fig. \ref{fig: Figure 4} shows the total operation costs under different uncertainties as the penetration level of RES power increases. The uncertainty level indicates how much the actual power output from RES differs from the forecasted amount. It measures the potential variation between the predicted power generation and the possible lower or higher actual outputs. This helps in understanding and managing the reliability of energy supply from RES. The figure shows that the operation costs decrease with increases in RESs penetration, showing economic benefits of using RESs. However, higher RES power penetration increases with uncertainties and requires higher reserves from diesel generators, resulting in higher costs. Considering RESs forecast reaches approximately 35\% of total demand, they can almost completely cover all the loads with power generation from the RESs, such that the operation cost stabilizes and much less power is produced by the diesel generator than its capacity.

\begin{figure}
    \centering
    \includegraphics[width=1\linewidth]{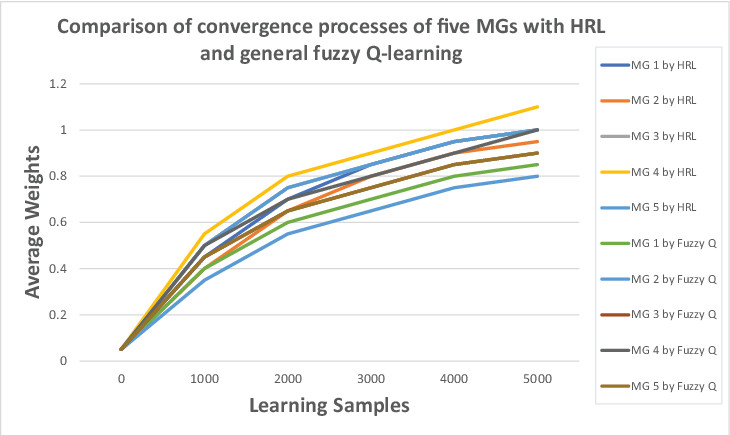}
    \caption{Convergence processes of five different types of MGs with HRL and general fuzzy Q-learning}
    \label{fig: Figure 5}
\end{figure}

Furthermore, considering the Distributed Online Economic Dispatch (D-
OED)  problem faced by the microgrids due to constraints and high dimension environment, the two-level HRL developed in~\cite{b87} to dispatch networked MGs online in a distributed manner is discussed. In contrast to the day-ahead dispatch, the online dispatch further optimizes the networked MGs with real-time data and eliminates prediction errors. The simulation results in Fig.~\ref{fig: Figure 5} show that the proposed HRL in~\cite{b87} outperforms the Fuzzy Q-learning and general online dispatch in supporting long-term operation, as the hierarchical learning framework needs far fewer training samples. The local operation of the proposed HRL can solve complex tasks, and each part of the system is controllable at runtime without manual intervention. The dispatch method proposed realizes distributed online self-adaptation, and the performance in cost reduction and stable operation has been highly improved. Ensuring the convergence of approximation parameters during the learning process is very important to achieve an accurate and stable system performance. When the iterative learning algorithms continuously adjust parameters to minimize errors to reach optimal values, it leads to convergence. When learning rates $\alpha_t,  \beta_t$ are time-varying, establishing general convergence conditions is necessary. The following are the general convergence conditions in the Greedy-Go Q-learning through which the estimated Q function converges to the unbiased estimation of optimal, Q~\cite{b87}:
\begin{enumerate}
\item  $0< \alpha_t < \beta_t < 1$
\item  $\sum_t \alpha_t = \sum_t \beta_t =1$
\item  $\sum_t \alpha_t^2 + \beta_t^2 < \infty$
\end{enumerate}
 where $\alpha_t >0$ and $\beta_t >0$  are necessary conditions and $\alpha_t < \beta_t$  is required to improve the efficiency of Greedy-Go algorithm. 
Time-varying learning rates help accelerating convergence and avoiding local minima, but challenges arise in maintaining stability and ensuring that the parameters converge properly. Choosing smaller values for $\alpha_t$ and  $\beta_t$ can ensure that the learning process converges successfully. However, while these smaller learning rates promote stability, they also slow down the speed at which the model learns. This means it will take longer for the model to reach optimal performance.
 Convergence processes of five different types of MGs with HRL and general fuzzy Q-learning during offline learning are compared in Fig.~\ref{fig: Figure 5}. As shown in the figure, in terms of speed and efficiency, HRL is better than fuzzy Q-learning. Because of the hierarchical structure of HRL, the dimensionality for learning reduces, and the learning efficiency improves through Principal Component Analysis (PCA).

\begin{figure}
    \centering
    \includegraphics[width=1\linewidth]{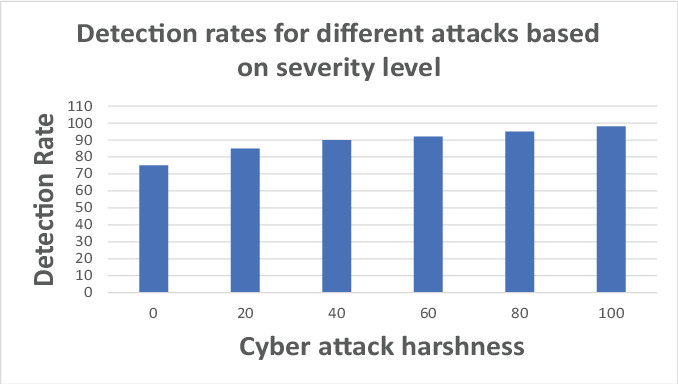}
    \caption{Detection rates for varying levels of attack severity using the proposed intrusion detection model in~\cite{b105} }
    \label{fig: Figure 6}
\end{figure}

The impact of a cyberattack on a microgrid depends on two main factors: the microgrid's operational mode (grid-connected and off-grid) and the attack's severity. In grid-connected mode, cyberattacks can lead to increased operational costs, higher power losses in distribution lines, and potential voltage instability~\cite{b110}. In off-grid mode, such attacks can have more severe consequences, including loss of power generation, imbalance between electricity supply and demand, and reduced power quality, which may lead to load shedding. Cyberattacks on microgrids can be categorized based on their impact and detection difficulty. The first category includes attacks that have immediate and significant effects, causing noticeable disruptions. These attacks are typically detected quickly due to their severity. The second category consists of more precise attacks that gradually affect the system over time~\cite{b111}. These are designed to remain undetected for extended periods, making them more difficult to identify and remove. To overcome the consequences of data integrity attacks, GANs were used in~\cite{b105}. In this method, two conflicting networks can be distinguished as: one generating false data and the other responsible for data classification. A new optimization technique has been proposed on a modified form of the Teaching–Learning-Based Optimization (TLBO) algorithm to reinforce the GAN model so that optimum features can be realized so as to improve the learning process. The results show the high performance of the proposed framework compared to well-known conventional detection frameworks, achieving a hit rate of 93.11\%, a miss rate of 6.89\%, a false alarm rate of 7.76\%, and a correct rejection rate of 92.24\%. For a detailed comparison of the model's performance, Fig.~ \ref{fig: Figure 6} provides the detection rate over different levels of cyberattack severity. The detection rate refers to the ratio of actual positive cases out of all detected cyberattack cases. As shown in the figure, for the hidden attack of data integrity, the detection rate remains at a lower value. However, with the increased severity levels of attacks, the detection rate is also very high. This performance is considered acceptable because if the level of attack severity is high, then the damage caused to microgrids is very harmful. It can even make them infeasible to run or even push them into off-grid operation. On the other hand, certain disturbances, which are not easily detected or are hidden mainly degrade the efficiency of power operation. Specifically, such hidden disruptions affect the generators, which may lead to reduced power output, increased fuel consumption, or unexpected wear and tear on the generators themselves.

\section{Challenges faced by AI-enabled microgrids}
While AI holds huge promise for revolutionizing microgrid energy management, many important issues need to be overcome before its full adoption. Addressing these challenges will pave the way for the widespread adoption of AI in microgrid energy management. This section discusses the main challenges facing the full integration of AI-based EMS in MGs. While a handful of opportunities and implementation challenges offered by this integration are summarized in Table 3, this section delves into the details of the potential challenges as discussed next.

\begin{table*} 
   \caption{Integration of AI in the EMS of MGs: Some prospects and open research issues}
\begin{center}
\begin{tabular}{|P{3.5cm}|P{1cm}|P{5.5cm}|P{4.5cm}|}
\hline
\textbf{AI technology} & \textbf{Ref.} & \textbf{Prospects} & \textbf{Implementation Challenges} \\
\hline
 Generative Adversarial Networks (GANs)& \cite{b84} 
 &Adopts a probabilistic model for RES output power prediction. Enhances Energy Efficiency and Demand Management & Uncertainties caused by load variations and component failures are not considered.  \\
\hline
 Radial basis function neural networks (RBF) Neural Networks& \cite{b85}& Optimizes the utilization of Renewable Energy Resources (RERs), minimizes emissions, and reduces production costs.& Degradation of battery and regulation of frequency and voltage is not considered. \\
\hline
 Recurrent Neural Network (RNN)& \cite{b86}& A solution to the optimal power flow problem for individual energy sources.&Complexity of the model is not considered. \\
\hline
Hierarchical Reinforcement Learning (HRL)& \cite{b87} & Enhance operation stability and reduce long-term operation costs.& Simulation-based model as opposed to implementation of the model in real-time.   \\
\hline
 Graph Convolutional Networks (GCN)& \cite{b88}& Proposes probabilistic power flow reducing computational time.&Simulation-based model as opposed to implementation of the model in real-time.  \\
\hline
 Feedforward Neural Networks & \cite{b90} & Improving the accuracy of short-term electricity demand forecasting.& Scalability. Applied only to small-scale distributed systems.\\

\hline
 Generative Adversarial Networks (GANs)& \cite{b105}& Examines the effect of attacks on data integrity for the central control of MGs.&Results are based on statistical model and heavily depend upon the data quality. Simulation-based only.\\
\hline

 Artificial Neural Networks (ANNs)& \cite{b106} & Power regulation within hybrid AC-DC distribution networks.&Applied only to the grid-connected mode of MG. \\
\hline
Convolutional Neural Networks and Gorilla Troop Optimization (CNN-GTO) & \cite{b92}& The proposed CNN-GTO models detected, classified, and
located feeder faults in the system with high accuracy&Applied and tested using only a software model of MG. \\
\hline
\end{tabular}
\label{tab3}
\end{center}
\end{table*}

\subsection{Data quality and availability}
AI technologies, especially in the context of microgrid energy management, rely heavily on the quality and quantity of data used for training and evaluation. However, a significant problem arises from the lack of genuine EMS data gathered from actual microgrid operations. This lack of authentic data has constrained researchers to mainly utilize simulated datasets in their research. This dependence on simulated data can have a detrimental effect on the accuracy of predictions and test results, eventually leading to inconsistencies in outcomes~\cite{b112}. The limited availability of real-world microgrid data arises from several factors. Firstly, because of the complex and sensitive nature of microgrid operations, it is important to have strict data privacy measures, making it challenging for researchers to obtain operational data. Secondly, the different types of microgrid configurations and operational protocols across different regions and applications further complicate the collection and standardization of data. Therefore, researchers are often forced to rely on simulated datasets that may not fully capture or represent the fine details and complexities of real-world microgrid scenarios.

The dependency on simulated datasets has consequential outcomes. Simulated data may fail to capture the randomness of renewable energy sources accurately, the ever-changing patterns of electricity demand, and the possibility of unforeseen disruptions in microgrid operations. These discrepancies between simulated data and real-world scenarios can result in deceptively positive assessments of AI performance, potentially hindering the advancement of robust and dependable microgrid energy management systems. To tackle this challenge, researchers and industry partners must collaborate on establishing secure and standardized data-sharing platforms that streamline the collection and distribution of real-world microgrid data. These platforms could incorporate de-identification techniques and data governance protocols to protect sensitive information while enabling researchers to access high-quality data for developing and evaluating AI-driven microgrid energy management systems. 

When data is incomplete or inaccurate, the considered problem can lead to less reliable predictions and suboptimal decisions by controllers, thereby reducing operational resilience. For example, the quality of data significantly affects the performance of forecasting models based on AI methods. Generally, data quality issues are addressed during the pre-processing phase before conducting predictive analytics. Many preprocessing techniques work under the assumption that missing data points occur occasionally. However, events such as communication failures, sensor malfunctions, or infrastructure damage from extreme incidents can cause prolonged periods of missing data, sometimes exceeding 15\% of the collected information~\cite{b113}. Current methods in the literature often handle these extensive gaps by ignoring them or filling them with zeros or average values, approaches that fail to represent the potential valuable information. Important characteristics to determine data quality are completeness, epistemic uncertainties, and drift.

When addressing missing data, it is essential to understand the underlying mechanisms before applying preprocessing methods like substituting missing values with column means or zeros. There are three primary types of missing data: Missing Completely at Random (MCAR), where the absence of data is entirely unrelated to any observed or unobserved values; Missing at Random (MAR), where the likelihood of missing data is related to observed variables but not the missing data itself; and Missing Not at Random (MNAR), where the missingness is directly related to the unobserved data~\cite{b114}. Recognizing these types is important for selecting the right method to handle missing data, ensuring more accurate and reliable analyses.
In~\cite{b115}, the authors examined grid-tied PV systems and identified that missing data in weather and energy generation often falls under the category of MCAR. To assess the extent of this missingness, they utilized Cohen's h, a statistical measure that quantifies the difference between two proportions. According to Cohen's distance measure, an h value of 0.2 indicates a small difference, 0.5 means a medium difference, and 0.8 represents a large difference. The formula for Cohen's $h$ is~\cite{b113}:
\begin{equation}
h = \frac{M_X - M_{\hat{X}}}{SD_{X\hat{X}}},        SD_{X\hat{X}} = \sqrt{\frac{SD_X^2 + SD_{\hat{X}}^2}{2}}
\end{equation}
where $M_X$ and $M_{\hat{X}}$ are the sample means of $X$ and $\hat{X}$, respectively, and $SD_X$ and $SD_{\hat{X}}$ are the respective standard deviations. This helps in understanding the impact of missing data on the performance of forecasting models. When a change in the probability distribution of a dataset, $X$ over time results in a new distribution $Y$, it is called as data drift. This can happen due to factors like seasonal variations, system changes (such as sensor wear or infrastructure damage), and adjustments in data calibration or equipment replacements. To measure data drift, the Minkowski distance is commonly used. By comparing the distances between observed and predicted values across datasets, data drift can be quantified. Another important metric is concept drift, which tracks changes in the relationship between input features $X_{PV}$ and target outputs $Y_{PV}$ over time~\cite{b114}. Unlike data drift, which can be measured using distance methods, concept drift is detected using residual error monitoring. When the error exceeds a certain threshold, it signals the need for model retraining or adjustment. Metrics like Normalized Root Mean Square Error (N-RMSE), Normalized Mean Square Error (N-MSE), Normalized Mean Absolute Percentage Error (N-MAPE), and both adjusted and unadjusted normalized R-square values are used to evaluate these residual errors and detect concept drift.

\subsection{Interoperability and integration}

The interoperability and executable integration in AI-enabled energy management for MGs need to incorporate seamless communication and compatibility amongst diverse systems. The integration of AI systems with infrastructure is complex due to the variety of devices, protocols, and software. In this context, standardized data exchange focuses on effectiveness in communication between the systems, which usually requires standard protocols and interfaces to be implemented~\cite{b116}. The IEEE 2030-2011 standard, titled ``Guide for Smart Grid Interoperability of Energy Technology and Information Technology Operation with the Electric Power System (EPS), End-Use Applications, and Loads," provides comprehensive guidelines for integrating energy and information technology systems within the smart grid framework ~\cite{b117}. Additionally, the IEEE 1547-2003 standard, known as the ``Standard for Interconnecting Distributed Resources with Electric Power Systems," sets criteria and requirements for connecting distributed energy resources, ensuring they work well and reliably within electric power systems ~\cite{b118}. Using these standards helps manage the complexities that come from having various devices, protocols, and software in microgrid infrastructures. However, many MGs still use older systems that may not support modern AI technologies, creating additional challenges. These traditional systems often cannot handle the large amounts of data and processing power that AI applications need, so they might require significant upgrades or complete replacements.

To bridge the gap between outdated and contemporary technologies, it is important to adopt interoperability standards like IEEE 2030 and IEEE 1547. These standards ensure unified operation and optimized energy management by providing a framework for integrating various systems and facilitating communication across various platforms.  In addition to the previously mentioned IEEE standards, several other protocols play a key role.  Communication protocol like IEC 61850 for intelligent electronic devices at electrical substations, enable efficient and reliable data exchange within the power system ~\cite{b119}. Another protocol called IEC 61970/61968 – Common Information Model (CIM) provides a common structure for energy management system data, facilitating interoperability between different systems and applications ~\cite{b120}. Adhering to these protocols not only improves compatibility but also promotes the scalability and resilience of AI-enabled energy management systems in microgrids.

\subsection{Scalability and operational issues}

One of the main challenges faced by future AI-based EMS is scalability. More specifically, as microgrid networks expand to include more renewable energy sources and energy storage components, and AI-powered energy management systems must be developed to handle the increasing complexity and uncertainty of such systems~\cite{b121}. Given that this ever-increasing data volume appears huge, AI solutions should be designed to manage this growth and not be hampered by scalability issues that often reduce performance. Besides, AI systems are dynamic—through periodic updates, proper maintenance, necessary amendments or corrections in the algorithms, upgradation of software, removal of software bugs, and better cybersecurity measures. This makes the process resource-intensive since both human and advanced technologies require investment regularly to ensure proper scalability and diligent maintenance. The AI systems should not become outdated or overburdened, undermining the purpose of optimizing microgrid operations.

\subsection{Standardization and regulatory issues}
 It is important to note that regulatory and policy challenges in integrating AI into microgrid energy management result from variations in compliance requirements across different regions in terms of regulations. In many cases, rules in the context of data privacy, ways of energy distribution, and even artificial intelligence from the side of different regions sometimes offer obstacles to applying a uniform solution to AI universally. For example, the lack of standard protocols on integration hinders developers; they do not have standards that ease considerations for interoperability and consistency in performance when used across platforms and systems. This lack of standardization may lead to disjointed efforts, high costs, and potential inefficiencies in deploying AI technologies at a large scale~\cite{b122}. Collective efforts on regulatory frameworks and standardized protocols would be necessary to bring out guidelines for adequate seamless integration of AI functionalities within such energy management systems in MGs.

\subsection{AI explainibility challenges}
In the context of AI-based EMS design and optimization, interpreting and explaining the behavior of AI models is another critical challenge. Challenges of explainability include a lack of understanding of the internal problem-solving process of the model and its outputs because the complexity of models is high and the `black box' paradigm adopted by many of the ML models. These challenges affect the accountability of information, and acceptance by users. AI systems are complex, especially in models like deep learning networks, whereby it may be challenging to understand the results since they function very much as black boxes, thereby making it challenging to understand which main inputs affect the ML performance or any form of bias in the output results ~\cite{b123}. While implementing ML algorithms into real-time microgrid energy management, it is important to have an overall understanding of the model's internal mechanism and decision-making processes, as various Explainable AI (XAI) methods have been recently published to address this issue. Despite these challenges, the potential benefits of AI in enhancing microgrid efficiency and reliability remain significant. 

There are many different XAI methods and tools available which are namely Local Interpretable Model-agnostic Explanations (LIME)~\cite{b124}, SHapley Additive exPlanation (SHAP)~\cite{b125}, GRADient Class Activation Mapping (GRAD-CAM)~\cite{b126}, and Deep Learning Important FeaTures (Deep-
LIFT)~\cite{b127}, Techniques like LIME and SHAP are versatile tools that can explain predictions from any machine learning model. They work by simplifying the model in a specific area, making it easier to understand how the model makes its predictions. Other methods such as GRAD-CAM and DeepLIFT are specifically designed for deep neural networks, providing insights by analyzing internal components like activation functions and weights. 

The authors in~\cite{b128} used a gradient boosting tree method to predict frequency stability indicators. To help operators better understand the model's predictions, they used the SHAP technique for explainability. In~\cite{b129,b130}the authors applied SHAP to a Deep Reinforcement Learning model to implement fair load shedding during undervoltage conditions. In~\cite{b131}, decision trees were used to classify whether a system's operating conditions were stable or unstable. The decision tree represented grid security rules, and the depth of the tree determined how interpretable the model was. This work balanced model accuracy and interpretability by using a modified optimal classification tree.

In the literature, XAI techniques were also used for identification of various fault scenarios in power grids. For example, in~\cite{b132}, grid disturbance events like generator tripping, line tripping, system oscillations, islanding, and load shedding were identified using a Long Short-Term Memory (LSTM) model. The deep SHAP method, which combines DeepLIFT and SHAP, was used to explain the LSTM model. The study also explored insights into misclassifications, which is a critical issue when using AI models in real-world applications. In~\cite{b133}, an XAI technique was applied to a long-term prediction model for forecasting cooling energy consumption. SHAP provided valuable insights that traditional black-box models could not offer. Similarly, in~\cite{b134}, an explainable long-term prediction model was introduced to study annual energy performance predictions using explainable methods.

One of the biggest challenges is to have models that are both highly accurate and easy to understand~\cite{b135}. While XAI methods can help make ML models more understandable, there are still issues to address, like creating standards and ensuring security. Using XAI with DL models is particularly difficult because, although DL models are very accurate, they are not transparent and are often used as 'black boxes'. This lack of transparency makes it hard to trust them without clear explanations.

\subsection{AI generalization and bias challenges}
The application of AI in the energy management of microgrids also faces challenges related to generalization and insufficient prevention of bias. In general, generalization capability allows AI models to function well with new data that is not included in the training data set; however, overfitting, inadequate training data, and the uncertainity in microgrid settings might affect the output~\cite{b136}. Bias in AI can lead to algorithm distortion and underrepresentation of some data formats, which leads to unfair or substandard decisions~\cite{b137}. 

Bias in AI models for microgrid energy management can lead to suboptimal performance when applied to different settings. For example, if an AI system is mostly trained on data from urban microgrids, it might not work well for rural or remote microgrids. These areas often have different energy usage patterns and resource availability, which can lead to inaccurate predictions or poor energy distribution decisions ~\cite{b138}. Similarly, if an AI model is trained mainly on data from sunny weather, it might struggle to predict energy needs accurately during cloudy or rainy days, causing energy shortages or overproduction. Additionally, AI models built using data from regions with advanced technology might not work in areas with limited or outdated infrastructure. For example, a microgrid system designed for places with strong internet connectivity might not function properly in remote areas with poor connectivity, resulting in unreliable energy management~\cite{b138}.

Some of the approaches to addressing these problems include working with diverse and inclusive datasets, regularly re-training models on new data for systematic validation and tests, using bias detection and reduction methods, and building understandable AI. By building models that generalize and are bias-free, the effectiveness, accuracy, and fairness in the distribution of energy to different loads may be improved in MGs.

\section{Future research directions and opportunities}
The fast changing advancements in AI tools and techniques are paving the way for many applications that are progressive in the field of microgrid EMS. As AI continues to evolve, one can expect to witness the emergence of innovative solutions that change the way microgrids operate, progress their resilience, and promote sustainable energy practices. Addressing the remaining challenges through future research and development will pave the way for the widespread adoption of AI technologies in microgrids. 

\subsection{Development of self-healing microgrids}
One of the most promising approaches for AI integration in microgrids lies in the development of self-healing microgrids. These intelligent systems would possess the ability to detect autonomously, isolate autonomously, and repair faults, enabling them to recover from disruptions and maintain a reliable supply of electricity without the need for human intervention~\cite{b139}. By utilizing sophisticated monitoring techniques and advanced AI algorithms, self-healing microgrids can pinpoint the location of faults, reroute power flows, and activate self-healing mechanisms within faulty components, ensuring uninterrupted operation and minimizing load-shedding time.

\subsection{Integration with blockchain technology}
Another application of AI in microgrids is the integration with blockchain technology which can transform energy transactions. Blockchain is an open distributed database that allows the creation of a permanent and easily identifiable record of multiple transactions between various parties and in various locations by maintaining its own replicated database of transactions~\cite{b140}. Blockchain's characteristics of immutability, transparency, and security make it an ideal platform for secure and auditable energy trading. By combining AI's capabilities of analysis with blockchain's secure accounting, one can promise a future where microgrids efficiently involve peer-to-peer energy transactions, thereby optimizing energy distribution and promoting a decentralized energy marketplace.

\subsection{Use of Internet of Things (IoT) }
The use of IoT presents yet another exciting opportunity for AI to optimize microgrid operations. IoT is defined as a global network of embedded sensors, software, and other related technologies that link physical objects to the Internet~\cite{b141}. These devices range from simple everyday appliances to complex industrial instruments, and connecting them creates a vast interconnected system. IoT devices, embedded within various microgrid components, continuously gather vast amounts of data, providing a comprehensive view of the microgrid's state and performance. AI algorithms can harness the vast amounts of data continuously gathered by IoT devices embedded within various microgrid components to analyze energy consumption patterns, predict renewable energy generation, and identify potential operational inefficiencies. By utilizing IoT-generated data, AI can optimize energy management strategies, enhance demand-side management programs, and proactively address maintenance issues, leading to improved microgrid efficiency and reduced operational costs.

\subsection{Addressing interpretability, data privacy, and scalability }

Research and development in the future should also consider concerns related to the interpretability of algorithms, data privacy and data quality and developing innovative solutions. These attempts in enhancement will pave the way for the extensive and efficient integration of AI technologies within microgrids. Additionally, future research on AI-based EMS should expand to include larger-scale, application-specific microgrids, such as networked and mobile microgrids. With the ongoing advancement of AI, one can expect a surge of innovative applications that will significantly elevate the capabilities of AI-driven EMS in microgrids, paving the way for a more robust, sustainable, and decentralized energy future.

\subsection{Roadmap for addressing challenges}
A structured roadmap is essential to effectively address the challenges identified in AI-enabled microgrids. Researchers and industry experts should work together to create secure, standardized platforms for sharing data. This ensures better data quality and availability for AI models. Investing in advanced sensors and developing efficient methods to fill in missing data, are bound to ensure the reliability of AI models, i.e., following established protocols like IEEE 2030-2011 and IEEE 1547-2003 to ensure microgrid components work well together~\cite{b116,b117}. Upgrading old systems and using modern communication standards like IEC 61850 and IEC 61970/61968 for smooth and efficient data exchange will help addressing challenges related to standardization~\cite{b119}. To address operational and scalability issues, AI systems must be designed that can manage huge amounts of data and should be updated regularly to keep up with growing demands. By continuously training staff and investing in new technologies, organizations can keep pace with the fast-evolving nature of AI systems. Developing unified regulatory frameworks and establishing standard protocols for AI integration can help address standardization and regulatory challenges. By using techniques like LIME and SHAP, AI decisions can be made easier to understand, ensuring a balance between accuracy, transparency, and security~\cite{b124,b125}. Finally, to reduce AI generalization and bias challenges, it is essential to employ diverse training datasets, implement bias detection mechanisms, and regularly train models. By following this proposed roadmap, the reliability, efficiency, and fairness of EMS in microgrids can be improved.

\subsection{Use of hybrid AI approaches}
Using hybrid AI approaches in microgrid EMS is a promising way to improve efficiency and resilience. Hybrid AI combines different AI methods to take advantage of their strengths, helping to handle the complex and ever-changing nature of microgrid operations. One effective hybrid approach is combining machine learning algorithms with optimization techniques. For example, combining neural networks with evolutionary algorithms can improve the accuracy of energy demand predictions while optimizing how resources are used~\cite{b142}. This helps the EMS adapt to changes in energy generation and consumption, keeping the grid stable. Another powerful technique is to combine deep learning models with reinforcement learning. Deep reinforcement learning enables real-time decision-making in the microgrid, such as adjusting energy prices or managing load shedding, by learning the best actions through continuous interaction with the system~\cite{b14}. This allows the EMS to react quickly to unexpected events, like equipment failures or sudden increases in demand, making the microgrid more resilient.

\subsection{Use of generative AI in future EMS}

The recent emergence of generative AI models is expected to bring in a cornucopia of innovation frontiers, especially in the context of EMS of microgrids, due to generative AI's better capability of analyzing large-scale data in light of the advents in large language models (LLM). Unlike the traditional AI methods, GANs and variational autoencoders (VAEs) are capable of producing synthetic data and, thus, offer better training, even when past data availability is scarce~\cite{b143}. This capability is especially important for forecasting energy demand, generating renewable energy, and market predictions where the procurement of complex data distribution is highly important for accuracy.  By using generative AI, one can infer about various possible decision-making situations and their results – for instance, various patterns of demand, extreme weather conditions, etc., which would enable microgrid operators to allocate resources more efficiently and prevent disruptions more effectively. For example in~\cite{b144}, Conditional Least Squares Generative Adversarial Network (C-LSGANs) was used to model the unpredictability of renewable energy sources (like solar or wind) in microgrids. These networks create realistic scenarios of how much energy renewables might produce under different conditions. This helps microgrid operators plan better energy distribution strategies, making the system more resilient and adaptable to changes in energy supply.

Moreover, the use of generative AI models promises to help optimizing future EMS platforms. More specifically, due to their ability to create a myriad of potential future scenarios, generative AI models can be used to improve the charging/discharging control of the storage system, the power load control, and the dispatch control to improve flexibility, grid monitoring and protection, and response to random changes. For example, the proposed method in~\cite{b145} uses real-time data from the grid (called point-on-wave measurements) along with generative AI to improve monitoring. This helps detecting faults instantly and gives microgrid operators a clearer picture of what is happening in the system. As a result, energy management becomes more reliable and secure. GANs were also used in~\cite{b105} to secure microgrid optimal energy management against data integrity attacks, as discussed in detail in the previous section.  Generative AI would also allow the development of innovative customer energy management plans for individual customers, thus achieving high customer satisfaction and maximum efficiency. In summary, in the context of future EMS, the utility of the generative AI approach can be well explored for generating new data and simulating various conditions, thereby adding significant value to the enhancement future of microgrid energy management systems’ functionality and reliability.

\section{Conclusion}
Many researchers are currently exploring how AI can be integrated into the EMS of microgrids. This paper highlights the key challenges and exciting opportunities associated with this integration. It emphasizes that AI technologies can support the widespread use of renewable energy sources and improve microgrid operations, playing a significant role in achieving carbon emission reduction goals by 2050. The paper demonstrates how AI can enhance energy efficiency, manage demand, lower operational costs, improve forecasting and predictive maintenance, and strengthen microgrid resilience and cybersecurity. Additionally, it identifies important research areas that need further exploration to fully unlock the benefits of AI-based EMS for microgrids. Overcoming these challenges through future research and development will enable the broader adoption of AI technologies, with transformative tools like generative AI, blockchain, and IoT set to revolutionize microgrid operations and their integration with large-scale power grids.



\sloppy

\fussy
\end{document}